\input harvmac

\def\p{\partial}

\def\half{{1\over 2}}

\def\x{\hat{x}}
\def\hp{\hat{p}}
\def\t{\hat{t}}

\Title{hep-th/0103107}{\vbox{\centerline{Dimensional Reduction via 
Noncommutative}
\vskip15pt
\centerline{Spacetime: Bootstrap and Holography}
}}
\vskip20pt

\centerline{Miao Li}
\vskip 10pt
\centerline{\it Institute of Theoretical Physics}
\centerline{\it Academia Sinica}
\centerline{\it Beijing 100080} 
\centerline { and}
\centerline{\it Department of Physics}
\centerline{\it National Taiwan University}
\centerline{\it Taipei 106, Taiwan}
\centerline{\tt mli@phys.ntu.edu.tw}

\bigskip

Unlike noncommutative space, when space and time are noncommutative, 
it seems necessary to modify the usual scheme of quantum
mechanics. We propose in this paper a simple generalization of the
time evolution equation in quantum mechanics to incorporate the feature
of a noncommutative spacetime. This equation is much more constraining 
than the usual Schr\"odinger equation in that the spatial dimension noncommuting 
with time is effectively reduced to a point in low energy. We thus call 
the new evolution equation the spacetime bootstrap equation, the 
dimensional reduction called for by this evolution seems close to what 
is required by the holographic principle. We will discuss
several examples to demonstrate this point.

\Date{March 2001}

\nref\ty{T. Yoneya, p. 419 
in ``Wandering in the Fields", eds. K. Kawarabayashi and
A. Ukawa (World Scientific, 1987) ;
see also p. 23 in ``Quantum String Theory", eds. N. Kawamoto
and T. Kugo (Springer, 1988).}
\nref\tamiaki{T. Yoneya, Mod. Phys. Lett. {\bf A4}, 1587(1989).}
\nref\ly{M. Li and T. Yoneya, hep-th/9611072, Phys. Rev. Lett. 78 (1997)
1219.}
\nref\lyrev{M. Li and T. Yoneya, ``Short-distance Space-time Structure
and Black Holes in String Theory: A Short Review of the Present
Status, hep-th/9806240, Jour. Chaos, Solitons and Fractals (1999);
for a more recent and elegant exposition, see T. Yoneya, 
``String Theory and the Space-Time Uncertainty Principle", hep-th/0004074, 
Prog.Theor.Phys. 103 (2000) 1081.}
\nref\transp{'t Hooft, Nucl. Phys. B256 (1985) 727; Int. J. Mod. Phys. A11 
(1996) 4623.}
\nref\ls{L. Susskind, Phys. Rev. {\bf D49}, 
6606(1994).}
\nref\miao{M. Li, ``Black Holes and Spacetime Physics in String/M Theory",
hep-th/0006024.}
\nref\holo{G. 't Hooft, gr-qc/9310026; L. Susskind, hep-th/9409089, J. Math. 
Phys. 36 (1995) 6377.}
\nref\mald{J. Maldacena, hep-th/9711200, Adv. Theor. Math. Phys. 2 (1998) 231.}
\nref\hl{P. M. Ho and M. Li, ``Large N Expansion From Fuzzy $AdS_2$",
hep-th/0005268, Nucl. Phys. B590 (2000) 198.}
\nref\stnon{N. Seiberg, L. Susskind and N. Toumbas, ``Strings in Background 
Electric Field, Space/Time Noncommutativity and A New Noncritical String Theory",
~hep-th/0005040, JHEP 0006 (2000) 021; R. Gopakumar, J. Maldacena, S. Minwalla
and A. Strominger, ``S-Duality and Noncommutative Gauge Theory",
hep-th/0005048, JHEP 0006 (2000) 036; J.L.F. Barbon and E. Rabinovici, 
``Stringy Fuzziness as the Custodian of Time-Space Noncommutativity",
hep-th/0005073, Phys. Lett. B486 (2000) 202.}
\nref\gm{J. Gomis and T. Mehen, ``Space-Time Noncommutative Field Theories 
And Unitarity", hep-th/0005129, Nucl. Phys. B591 (2000) 265.}
\nref\np{V. P. Nair and A. P. Polychronakos, ``Quantum mechanics on the 
noncommutative plane and sphere", hep-th/0011172.}
\nref\gkl{J. Gomis, K. Kamimura and J. Llosa, ``Hamiltonian Formalism 
for Space-time Non-commutative Theories", hep-th/0006235, 
Phys. Rev. D63 (2001) 045003.}

String theory intrinsically exhibits spacetime uncertainty, as first pointed out
in \refs{\ty,\tamiaki} and later examined in some nonperturbative context in 
\refs{\ly,\lyrev}. This seems to indicate that space and time will become 
noncommutative in a more fundamental formulation of string/M theory. On the other 
hand,  quantum black hole physics also cries for noncommutative spacetime
\refs{\transp,\ls}, for a different but somewhat related reason. 
For a recent review on the two aspects of spacetime uncertainty, see \miao.
These observations
recently led many people to believe that in string theory and more generally
in a theory of quantum gravity, the notion of spacetime will become
approximate, and that a fundamental formulation of string/M theory should
unify kinematics and dynamics of spacetime into a single framework.

A noncommutative spacetime implies that some dimensions are redundant in
a quantum mechanical description, much as in quantum mechanics information
is required of the wave function only for one variable in a pair of conjugate 
variables. This, we believe, is closely related to the holographic principle in
a quantum gravity theory \holo\ in which a few spatial dimensions are reduced.
Thus, it is very likely that in a theory
exhibiting holography, such as the AdS/CFT correspondence \mald, space and time
are noncommutative. The first concrete proposal in string theory, to
the best of our knowledge, appears in \hl\ in which a fuzzy $AdS_2$ is proposed
to describe the quantum gravity effects. Indeed, a correct pattern of $1/N$
expansion as well as of nonperturbative effects in large N emerges, the latter
can be interpreted as caused by the instanton processes of fragmentation of smaller
$AdS_2$.

It is possible to construct theories with noncommutative spacetime by switching
on an electric field on D-branes. Unlike in the magnetic case, no
decoupling limit exists for one to obtain an effective field theory
\stnon. If one tries to define a field theory on a noncommutative spacetime
perturbatively using Feynman rules, one runs into contradiction: The theory
would be nonunitary perturbatively \gm. Even if such inconsistency was not
there, one would still have problem to define such a theory nonperturbatively.
To define a quantum amplitude between two field configurations at two
given times, one would have to specify many time derivatives of fields, since
the action contains infinitely many time derivatives. This in a way
requires one to introduce many more degrees of freedom than those the theory
initially contains.

We propose to study noncommutative spacetime in the opposite
direction, namely instead of introducing more degrees of freedom we shall
try to reduce the number of degrees of freedom, as we explained above
that interesting physics demands such a procedure. To this end, it appears to 
us that it is necessary to reformulate quantum mechanics when time is regarded 
as an operator rather than a continuous evolution parameter.
As we shall explain shortly, usual quantum mechanics for a single degree of
freedom can be easily formulated on a noncommutative space, and the second
quantization, namely a field theory, can also be formulated. However, 
one can not write down a Schr\"odinger equation in which noncommutativity of
space and time manifests, this is simply because in the such an evolution
equation time is always treated as a c-number variable. We will propose a 
simple generalization of the time evolution equation. This equation departs
from quantum mechanics in a minimal fashion, and is much more constraining
than the Schr\"odinger equation, thus enabling dimensional reduction.
If the system under study also involves other commuting spatial dimensions,
we will be led to an effective low dimensional theory defined on the commuting
space, in the low energy limit. In the dimensional reduced theory, usual
quantum mechanics is recovered.

\noindent $\bullet$ {\it Noncommutative space}

We start with the problem of formulating quantum mechanics on a noncommutative
plane. This subject has not received much attention except \np, the reason
is probably that for many experts it is rather straightforward to do so.
We present a discussion here for the purpose to distinguish this case from
our subject of main interest: noncommutative spacetime. Our discussion
is somewhat different from that in \np. Let us start with the following
operator algebra
\eqn\opa{\eqalign{[\x^i, \x^j]&=i\epsilon^{ij}\theta, \quad
[\x^i,\hp_j]=i\delta_{ij},\cr
[\hp_i,\hp_j]&=0.}}
Since $\x^i$ are noncommuting, they do not have mutual eigenstates in the
Hilbert space. $\hp_i$ however are still commuting, they do have mutual
eigenstates
\eqn\meigen{\hp_i|p\rangle =p_i|p\rangle,}
or for a state $|\Psi\rangle$ expanded in the above basis
$\Psi (p)=\langle p|\Psi\rangle$:
\eqn\prep{\hp_i\Psi(p)=p_i\Psi(p).}
The multiplication on the R.H.S. is the usual one.

In the momentum representation, the position operators are realized by
\eqn\preal{\x^i=i\p_{p_i}-{\theta\over 2}\epsilon^{ij}p_j,}
when acting on $\Psi(p)$. Now although it is impossible to define
the eigenstates of operators $\x^i$, one can define the state
$|x\rangle$ by $\langle x|p\rangle =\exp (ip_ix^i)$ so that
\eqn\conwf{\x^i\Psi (x)=x_i\Psi(x)+{i\theta\over 2}\epsilon^{ij}
\p_j\Psi(x)=x_i*\Psi(x),}
where we used the standard $*$-product. 

It is not hard to convince oneself that for any operator as a
function of $\x^i$, if Weyl ordered, its action on the wave function
$\Psi (x)$ is given by the $*$-product between itself and the wave
function. If there is a potential for a single particle depending
only on $\x$, $V(\x)$, then one has $V(\x)\Psi(x)=V(x)*\Psi(x)$. 
The Schr\"odinger equation can be written as
\eqn\scheq{i\p_t \Psi (t,x)=(-{1\over 2m}\p_i^2 +V(x)*)\Psi(t,x).}
Apparently, there are just as many solutions to the above
equation as many states in the Hilbert space, and the latter
is the space of a certain class of functions of $x$. This 
means that although the plane is noncommutative, there is
no dimensional reduction at all.

In the second quantization, both time and space coordinates become
dummy variables, and it is the field itself acts as an operator. 
For a nonrelativistic particle moving on a noncommutative plane and
under a potential, the Hamiltonian is written as
\eqn\sech{H=\int d^2x\left(-{1\over 2m}\overline{\Psi}(x)\p_i^2
\Psi(x)+\overline{\Psi}(x)*V(x)*\Psi(x)\right).}
If the interaction is a type of self-interaction, one replaces
$V(x)$ in \sech\ by an operator such as $\overline{\Psi}(x)*
\Psi(x)$, one ends up with a noncommutative field theory.

\noindent $\bullet$ {\it Noncommutative spacetime} 

We have seen that all is fine with a noncommutative space. The
standard quantum mechanics applies rather straightforwardly. 
We now postulate that for some reason time becomes noncommutative
with a spatial dimension $x$. We have an operator algebra
\eqn\stal{[\t, \x]=i\theta,\quad [\x,\hp]=i.}
If we set the speed of light $c=1$, $\theta$ has a dimension
of length square. Apparently, there is trouble with the
usual idea about the Hilbert space. Assume that the Hilbert space
is the usual representation of the Heisenberg algebra
$[\x,\hp]=i$, then since $[\x, \t+\theta\hp]=0$, one
deduces
\eqn\const{\t=-\theta\hp+f(\x),}
that is, time is no longer an independent dummy variable,
and is proportional to momentum. This in itself is correct
for many physics phenomena, but impedes generalization of
the Schr\"odinger equation. For now the standard wave function
$\Psi(t,x)$ would become a function of the form
$\Psi(p,x)$. This does not make sense for $\Psi$ as a wave
function. Only the Wigner distribution depends both
on $x$ and $p$. If on the other hand one insists on treating
the wave function as a function of only one variable,
then in the standard Schr\"odinger equation the commutator
$[\t,\x]=i\theta$ plays no role at all.

Thus it seems necessary to go beyond quantum mechanics, as
first emphasized in the conclusion section of \miao. We will
still postulate that it makes sense to talk about the
joint function $\Psi (t,x)$, but one has to generalize
the Schr\"odinger equation. The right version of the
time evolution equation to be generalized seems to be the 
integrated form
\eqn\isch{\Psi(t,x)=e^{-itH(\hp,\x)}\Psi(x).}
Now for {\it any} initial wave function, there is no
problem in using the above
equation to evolve it to time $t$. Due to the simple fact 
that
\eqn\comp{e^{-it'H}e^{-itH}=e^{-i(t+t')H},}
one has the composition law
\eqn\compl{e^{-it'H}\Psi(t,x)=\Psi(t+t',x),}
as demanded by the requirement that the fundamental
evolution equation must be time-translationally invariant.

Now when $\t$ is noncommuting with $\x$, one might be
motivated to generalize \isch\ to the form
\eqn\ngen{\Psi (t,x)=e^{-itH(\hp,x)}*\Psi(x),}
where $\Psi(x)$ is the wave function at $t=0$. The
above $*$-product is a generalized one defined by
\eqn\gstar{f(t',x)*g(t,y)=e^{{i\theta\over 2}
(\p_{t'}\p_y-\p_t\p_x)}f(t',x)g(t,y).}
Such a generalized $*$-product for functions at
different points have been extensively used recently
in constructing gauge invariant observables in a
noncommutative Yang-Mills theory. Although in \ngen\
the two times are different, the space points in
two functions are the same. Also we implicitly assume
in \ngen\ that $\hp$ is identified with $-i\p_x$,
and in the $*$-product it is treated as a dummy
variable.

Evolution equation \ngen\ differs from the quantum mechanical
one \isch\ in that we need to specify all the
time derivatives of $\Psi(x)$ at the initial time
$t=0$. This at the first sight requires introducing
more degrees of freedom. We shall see shortly that
the opposite is true. The evolution form is not
guaranteed to be time-translationally invariant,
since
\eqn\ncomp{e^{-it'H(\hp,x)}*e^{-itH(\hp,x)}\ne 
e^{-i(t+t')H(\hp,x)}.}
Thus, instead of working with \ngen\ we propose to
work with the following equation
\eqn\boots{\Psi(t'+t,x)=e^{-it'H(\hp,x)}*\Psi(t,x).}
The above equation is required to be valid for
arbitrary $t$ and $t'$, therefore by definition the
composition law is satisfied. \boots\ differs from \ngen\
in that on the R.H.S. of \ngen\ $\Psi$ and
its time derivatives can be arbitray at $t=0$, while
\boots\ is better viewed as a contraining equation rather
than an evolution equation, since one can not arbitrarily
specify $\Psi$ and its time derivatives at $t$, their
form must be consistent with the function on the L.H.S.
We shall see soon that indeed this equation will
drastically reduce the number of solutions. If one
views \boots\ both as an evolution equation as well
as a constraining equation, it is better called the
bootstrap equation, in the sense that time and space
bootstrap themselves.

By the definition of the generalized $*$-prouduct,
we have the following algebra
\eqn\stgal{[t,x]=[t',x]=i\theta \quad 
[t,t']=0,}
where we omitted hat for all variables. This algebra
is certainly consistent with the one we started with in
\stal. We now proceed to examine the implications of our
main equation \boots.

\noindent $\bullet$ {\it H(p,x)=H(p)}

The simplest Hamiltonian to work with is a Hamiltonian
depending only on momentum. In this case,
\boots\ has the form
\eqn\expan{\Psi(t+t',x)=e^{-it'H(p)}\Psi(t,x)
+\sum_{n\ge 1}{1\over n!}({\theta H\over 2})^ne^{-it'H}
\p_x^n\Psi(t,x).}
At $t'=0$, we deduce
\eqn\constr{\sum_{n\ge 1}{1\over n!}({\theta H\over 2})^n
\p_x^n\Psi(t,x)=0.}
Once this constraining equation is satisfied, the
sum in \expan\ automatically vanishes, and we have
\eqn\qmev{\Psi(t+t',x)=e^{-it'H(p)}\Psi(t,x),}
the standard quantum mechanical evolution equation.
Both the constraining equation \constr\ and \qmev\
are linear, thus an arbitrary solution can be decomposed
into a complete basis of solutions. Any solution in
this basis satisfies the composition law, since \qmev\
is the usual quantum mechanical equation.

Without loss of any information, we will set $t=0$
in \constr\ and proceed to solve it. Denote the
wave function at $t=0$ by $\Psi(x)$. Using Fourier
transformation
\eqn\fouri{\Psi(x)=\int dp \tilde{\Psi}(p)e^{ipx},}
we find
\eqn\inte{\int \left(e^{i{\theta\over 2}pH(p)}-1\right)
\tilde{\Psi}(p)e^{ipx}dp=0.}
So the constraining equation boils down to the simple
equation
\eqn\sime{\left(e^{i{\theta\over 2}pH(p)}-1\right)
\tilde{\Psi}(p)=0.}

For $H(p)$ non-singular at $p=0$, there is
a obvious solution $\Psi (p)=\delta (p)$. If $H(p)$
also vanishes at $p=0$, there are more solutions.
For instance for a ``nonrelativistic" particle
$H(p)=p^2/(2m)$, there are three solutions localized
at $p=0$:
\eqn\lesol{\Psi (p)=a\delta(p)+b\delta'(p)
+c\delta''(p).}
In the position space, we have
\eqn\pssol{\Psi(x)=a+bx+cx^2.}
After applying the evolution equation \qmev, we get
\eqn\tisol{\Psi(x)=a+bx+c(x^2+{it\over m}).}
It is easy to check that indeed this family of solutions
satisfies the bootstrap equation \boots. The energy of
these solutions is zero. We conclude that in the
low energy sector, the number of solutions is finite.
If $H(p)\sim p^m$, there are $m+1$ solutions in
the zero-momentum sector.

For nonvanishing $p$, the factor in the parenthesis
in \sime\ vanishes if
\eqn\qasol{p_nH(p_n)={4\pi n\over\theta},}
so there is a solution $\Psi(p)\sim\delta (p-p_n)$.
The momentum is quantized, although we start with a 
noncompact dimension $x$! The simplest case is when
$H$ is a constant ${\cal E}$, 
\eqn\qumom{p_n={4\pi n\over \theta{\cal E}}.}
This formula indicates that the particle effectively
moves on a circle of radius
\eqn\rad{R={\theta{\cal E}\over 4\pi}.}
The smaller the parameter $\theta$ is, the smaller
the circle, the larger the momentum, while the
energy is always a constant. Since in string theory
$\theta$ is related either to the string scale
or the Planck scale, the nonzero momentum is
actually quite high, one might say that the
low momentum sector ($p$=0) is one dimensional.
Even when one takes all nozero momentum \qumom\
into account, the original quantum mechanical
Hilbert space is reduced to one in which momentum
is discrete. We see that indeed we achieved what
we hoped: There is a drastic reduction of degrees
of freedom.

The next interesting case is when $H(p)=|p|$.
We find
\eqn\mqu{p_n^2={4\pi n\over\theta}.}
This quantization condition is similar to the
string spectrum \foot{I am grateful to P. M. Ho
for suggesting this analogue.}. In this case,
the energy is no longer constant, but proportional
to $\sqrt{n/\theta}$. Again, for small $\theta$,
all these are in the high energy sector. There are
two states in the low energy sector.

If we have a ``nonrelativistic" particle, we have
\eqn\nrsol{p_n^3={8\pi m n\over\theta}.}
In this case there is no direct analogue in familar
models. 

We notice that as long as $\theta\ne 0$, the
nonvanishing momentum is always quantized. This
phenomenon has no smooth limit at $\theta =0$,
where the constraining equation \constr\ is
always satisfied for any $p$.

Suppose in addition to $x$, there are more spatial
dimensions and these are always commuting with time.
Assume $H=H(p,q)$, $q$ is the momentum in the commuting
directions. The quantization condition \qasol\ is still
valid. In this case $p_n$ depends continuously on
$q$, in general we have $p_n=p(n,q)$. $q$ for a given
$n$ can vary in a range continuously. For example, if
\eqn\rehal{H=\sqrt{p^2+q^2},}
then
\eqn\consol{p_n^2=\half\left( (q^4+4({4\pi n\over\theta})^2
)^{1/2}-q^2\right).}
For small $q$, namely $q^2\ll 1/\theta$
\eqn\lowm{p_n^2\sim {4\pi |n|\over \theta}.}
The energy is also of this order. For large $q$,
that is $q^2\gg 1/\theta$, 
\eqn\himo{p_n\sim {4\pi |n|\over |q|\theta}.}
The energy is approximately $|q|$. In both
cases, the energy is no smaller than $1/\sqrt{\theta}$.

We are more interested in the low energy sector, namely
when $p=0$. For the above relativistic model, there
is always only one state if $q\ne 0$. In this
case $E=|q|$, the theory is completely reduced to
a zero mass relativistic particle in the commuting
space, and moreover the usual quantum mechanics
is valid for this dimensionally reduced theory.

One can proceed to discuss different Hamiltonians
when $q$ is involved. Suffices it to say that
in addition to the low energy sector which is
finite dimensional in terms of $x$-dimension, all
other solutions are highly energetic if $\theta$ is
very small. In all these theories, the
dimensional reduction occurs in the $p$-space.

\noindent $\bullet$ {\it H=H(x)}

Next we consider the case when the Hamiltonian
is a function of only $x$. In order to solve
the bootstrap equation \boots\ completely, we
need the following formula for the $*$-product,
\eqn\intrep{\eqalign{f(t',x)*g(t,x)&=
{1\over (\theta\pi)^2}
\int\prod_i dt_idx_if(t_1,x_1)g(t_2,x_2)\cr
&\exp\left({2i\over\theta}[(t'-t_1)(x-x_2)
-(t-t_2)(x-x_1)]\right).}}
For $f(t_1,x_1)=\exp(-it_1H(x_1))$ and
$f(t_2,x_2)=\Psi(t_2,x_2)$, one may integrate
out $t_1$ and then $x_2$ first to find the
bootstrap equation
\eqn\reev{\eqalign{\Psi(t+t',x)=&{1\over\theta\pi}
\int dt_2dx_1\Psi(t_2,x+{\theta\over 2}H(x_1))\cr
&\exp\left(-it'H(x_1)-{2i\over\theta}
(t-t_2)(x-x_1)\right).}}
Use the definition
\eqn\enr{\tilde{\Psi}(E,x)={1\over 2\pi}
\int dt e^{iEt}\Psi(t,x)}
to integrate out $t_2$
\eqn\sinin{\eqalign{\Psi(t+t',x)&=
{2\over\theta}\int dx_1\tilde{\Psi}({2\over\theta}
(x-x_1),x+{\theta\over 2}H(x_1))\cr
&\exp\left(-it'H(x_1)-{2i\over\theta}
t(x-x_1)\right).}}
Fourier transforming both sides of the above
equation with respect to $t$ results in
\eqn\funcre{e^{-it'E}\tilde{\Psi}(E,x)
=e^{-it'H(x-{\theta\over 2}E)}\tilde{\Psi}
(E, x+{\theta\over 2}H(x-{\theta\over 2}E)).}

The functional relation \funcre\ is easy to
solve. If $H$ is a nontrivial function of $x$,
the $t'$ dependence of the two sides of \funcre\
do not agree unless $\tilde{\Psi}(E,x)$ concentrates
at the point when $H(x-{\theta\over 2}E)=E$, that is
\eqn\deld{\tilde{\Psi}(E,x)\sim \delta (
H(x-{\theta\over 2}E)-E).}
Compare the $x$ dependence of the two sides of 
\funcre\ we find that whenever
\eqn\revi{H(x-{\theta\over 2}E)=E}
$x$ and $E$ must also satisfy
\eqn\frevi{H(x+{\theta\over 2}H(x-{\theta\over 2}E)-{\theta\over
2}E)=E.}
Substituting $H(x-{\theta\over 2}E)$ in the
above relation by $E$, we have
$$H(x)=E.$$
We conclude that the following two equations must
be satisfied simultaneously
\eqn\twof{H(x-{\theta\over 2}E)=E,\quad
H(x)=E.}

For a monotonic function $H(x)$, the two
equations in \twof\ can not possibly be
satisfied at the same time unless $E=0$. Thus,
we have
\eqn\fsol{\tilde{\Psi}(E, x)=f(x)\delta (H(x))
\delta (E).}
To determine the form of $f(x)$, substituting
the above ansatz back to \funcre\ we find
\eqn\ssol{f(x)\delta (H(x))=f(x+{\theta\over 2}
H(x))\delta(H(x+{\theta\over 2}H(x))).}
This equation does not always have a solution.
For instance if $H=\omega^2 x$, there is no
solution. It is easy to convince oneself that
any zero of $H(x)=0$ must be of higher
order. There is a solution when $H=\omega^2x^3$,
in this case $f(x)\sim x$. In other word,
\eqn\asol{\tilde{\Psi}(E,x)=\delta (E)\delta(x).}
The solution is completely localized at
$x=0$ and the energy vanishes there. This
reduction of degrees of freedom is even more
complete than in cases when $H$ is a function of
momentum $p$.

If $H(x)$ is not a monotonic function of $x$,
there can be more than one solutions. Consider
$H(x)=\omega^2x^2$.  The two equations of \twof\
can be solved if $E=0$ and $\tilde{\Psi}$ is
localized at $x=0$. These equations can also be solved
at the same time if 
\eqn\consol{E=({4\over\theta\omega})^2,}
and $\Psi$ is localized at
\eqn\loc{x={4\over\theta \omega^2}.}
However this solution is not good, as for the value
\consol, there are more zeros of \frevi\ than those of
\revi. In other words, to have nonvanishing energy 
in a soluton, we need a complicated function $H(x)$. 
As a more interesting example, take 
\eqn\pep{H(x)={\cal E}\sin (\omega x).}
This function has a period ${2\pi\over
\omega}$, so a class of solutions is
\eqn\acl{E_n={4\pi n\over\theta\omega},\quad
\sin (\omega x_n)={4\pi n\over \theta\omega
{\cal E}}.}
When $\theta\omega{\cal E}\gg 1$, there are
many such localized solutions.

Now if there are commuting spatial dimensions,
and $H$ depends also on either the position in
this space or associated momentum, or both, there can be 
solutions to the two equations of \twof. 
In this case, both $E$ and the localization
$x$ will depend on these extra parameters. If
$H$ is always a monotonic function of $x$,
no interesting solution will be obtained, since
$E$ always vanishes thus is independent of the
parameters in the commuting space. If $H$ is
not a monotonic function, one can have some fun
here. For instance, replace $\omega^2$  
in the Hamiltonian $H=\omega^2x^2$ by
$c\theta^{-2}|q|^{-1}$, where $q$ is the momentum along the
commuting space and $c$ is a dimensionless parameter, 
we have
\eqn\funny{E={16\over c}|q|,\quad
x={4\over c}\theta |q|.}
The first identity in the above is a 
relativistic dispersion relation, while the
second relation looks like the UV/IR relation
in the AdS/CFT correspondence. Of course
we can not possibly hope to get a realistic
model of AdS/CFT correspondence in this fashion
any time soon. The relations in \funny,
though encouraging, can not be taken literally as we remarked
before.

\noindent $\bullet$ {\it The general case}

We have seen that when $H$ is a function of $p$ only,
the wave functions are localized in the $p$ space;
when $H$ is a function of $x$ only, the wave
functions are localized in the $x$ space, and the
number of possible solutions is more reduced in this
case. We expect that more interesting phenomenon
will happen if $H$ is a function of both $p$ and
$x$. This problem can be hardly solved generally,
so we will be content with a remark.

Similar to \reev, for $H(p,x)$, the bootstrap
equation \boots\ can be written in an integral
form
\eqn\inteq{\Psi(t+t',x)=\int dt_2dx_2
G(t,t',x, t_2,x_2)\Psi(t_2,x_2).}
This in general is a difficult integral
equation to solve. For example, for $H$
to be the Hamiltonian of a harmonic oscillator,
we do not know how to solve this equation
at present. The kernel $G$ in no way
depends on $t$ and $t'$ in the combination
$t+t'$, so the composition law in time
is a rather strong constraint on the possible
wave functions. We get a projection equation
at $t'=0$, this requires that the kernel
$G(t,0,x,t_2,x_2)$ is a projection operator,
and in general it has only a finite dimensional
invariant subspace of the space of all functions. 
So the dimensional reduction is a general mechanism.

\noindent $\bullet$ {\it Conclusion and discussions}

First of all, we want to emphasize that there is
no issue of violating unitarity in the approach
we suggest here. In our view, as we explained in
the beginning of this paper, the usual action principle
for a field theory on a noncommutative spacetime
is an ill-defined notion. One may try to get rid
of the conceptual issues as well as the technical ones by
introducing more degrees of freedom, or going into
higher dimensions, as advocated in \gkl. This attempt
runs in the opposite direction to our philosophy.

The quick investigation of the idea about going
beyond quantum mechanics for a noncommutative 
spacetime in this paper leaves many problems unanswered.
On the technical side, one would like to solve
the bootstrap equation for more Hamiltonians
to gain experience and insight about how dimensional
reduction occurrs. The bootstrap equation proposed
here does not have a differential form, one that
would be close to the original Schr\"odinger equation.
The best one can hope for is to formulate a path integral
form, thus spelling out this procedure in an action
principle.

It is also interesting to see
whether the simplest form of the bootstrap equation
\boots\ has other variants. An important point is that
not only one wants to incorporate the noncommutative
spacetime, one also wants to reduce to quantum mechanics
after the dimensional reduction is achieved.
It therefore appears that not many such variants
exist.

To make contact with field theories and string theory,
ultimately one needs to study second quantization.
Now wave functions are promoted to operators,
these operators again should satisfy the bootstrap
equation when on-shell. Now a second quantized Hamiltonian
is to be constructed in such a way that the bootstrap
equation is the equation of motion resulting from
it. Or is there a second quantized Hamiltonian at all?

To study the AdS/CFT correspondence, in particular to 
understand how holography really works, there is
a long way to go. Apparently in the usual formulation
of AdS/CFT, the two sides have different dynamical variables
and Hamiltonians, although there is a dictionary between
them. As suggested in \hl, the noncommutative spacetime
is a notion for the bulk theory, so when applying the present
idea to that context, the resulting boundary theory may
well be different from the known CFT. It might be
that there is a set of bulk variables different from
those known in the bulk theory (more fundamental
variables in some sense?), such that when dimensionally
reduced, one gets the CFT. 

It has been advocated by some people for some time that 
in a fundamental formulation of the bulk variables
in string/M theory, there is a large gauge symmetry.
Once the symmetry is fixed, one is led to the boundary
theory just as what happens in Chern-Simons theory
with a boundary. This suggestion also demands to
construct new bulk variables, just as in our proposal.
We however view our proposal as a more attractive one
for several reasons: 

\noindent 1. Noncommutative spacetime by
now is commonly believed to be a general feature of
string/M theory, apparently it must be linked to
holography. Our proposal shows that indeed this is the
case. 

\noindent 2. In the bootstrap equation, quantum mechanics
plays an essentially role. Spacetime uncertainty
is a dynamical phenomenon, in the very beginning
we build this in the fundamental equations.
That AdS/CFT works precisely because on each side
quantum mechanics is taken into account. This goes without
saying on the CFT side. On the AdS side, for example,
the giant graviton phenomenon is a quantum mechanical, 
nonperturbative one.

\noindent 3. In our scheme we are beginning to see how
spacetime and quantum mechanics are intimately tied up.
Also, what is more exciting than going beyond quantum
mechanics?

\noindent 4. Finally, we have presented a concrete and
very simple scheme,
it would be wasteful if we do not explore it in depth.

Acknowledgments. 
The idea presented here occurred to the author while he participated the
M theory workshop at ITP, Santa Barbara, and visited the high energy
group of Univ. of Utah. He is grateful to the organizers of the M
theory workshop and the high energy group of Univ. of Utah, in
particular Y. S. Wu, for hospitality. Discussions with P. M. Ho
and Y. S. Wu are acknowledged.
This work was supported by a grant of NSC, and by a 
``Hundred People Project'' grant of Academia Sinica and an outstanding
young investigator award of NSF of China.

\vfill
\eject

\listrefs
\end